\documentclass[aps,twocolumn,groupedaddress,superscriptaddress,floatfix,prl]{revtex4-1}
\usepackage{amsmath}
\usepackage{amssymb}
\usepackage{graphicx}
\usepackage{textcomp}
\usepackage{bm}

\usepackage[usenames,dvipsnames]{color}	

\begin{document}

\title{An Artificial Neuron Implemented on an Actual Quantum Processor} 

\author{Francesco Tacchino}\email{francesco.tacchino01@ateneopv.it}    
\affiliation{Dipartimento di Fisica, Universit\`a di Pavia, via Bassi 6, I-27100, Pavia, Italy}
\author{Chiara Macchiavello}\email{chiara.macchiavello@unipv.it}
\affiliation{Dipartimento di Fisica, Universit\`a di Pavia, via Bassi 6, I-27100, Pavia, Italy}
\affiliation{INFN Sezione di Pavia, via Bassi 6, I-27100, Pavia, Italy}
\affiliation{CNR-INO, largo E.\ Fermi 6, I-50125, Firenze, Italy }
\author{Dario Gerace}\email{dario.gerace@unipv.it}
\affiliation{Dipartimento di Fisica, Universit\`a di Pavia, via Bassi 6, I-27100, Pavia, Italy}
\author{Daniele Bajoni}\email{daniele.bajoni@unipv.it}
\affiliation{Dipartimento di Ingegneria Industriale e dell'Informazione, Universit\`a di Pavia, via Ferrata 1, I-27100, Pavia, Italy}

\date{\today}
\begin{abstract} 
Artificial neural networks are the heart of machine learning algorithms and artificial intelligence protocols. Historically, the simplest implementation of an artificial neuron traces back to the classical Rosenblatt's ``perceptron'', but its long term practical applications may be hindered by the fast scaling up of computational complexity, especially relevant for the training of multilayered perceptron networks. Here we introduce a quantum information-based algorithm implementing the quantum computer version of a perceptron, which shows exponential advantage in encoding resources over alternative realizations. We experimentally test a few qubits version of this model on an actual small-scale quantum processor, which gives remarkably good answers against the expected results. We show that this quantum model of a perceptron can be used as an elementary nonlinear classifier of simple patterns, as a first step towards practical training of artificial quantum neural networks to be efficiently implemented on near-term quantum processing hardware. 
\end{abstract}

\maketitle

\section{Introduction}
\label{Introduction}

Artificial neural networks are a class of computational models that have proven to be highly successful at specific tasks like pattern recognition, image classification, and decision
making \cite{Schmidhuber2015}. They are essentially made of a set of nodes, or neurons, and the corresponding set of mutual connections, whose architecture is naturally inspired by neural nodes and synaptic connections in biological systems \cite{Schmidhuber2015,Zurada:intro_ANN_1992}. In practical applications, artificial neural networks are mostly run as classical algorithms on conventional computers, but considerable interest has also been devoted to \emph{physical} neural networks, i.e. neural networks implemented on dedicated hardware~\cite{Rojas_ANN_Introduction,Schuman_survey_neuromorphic_arxiv2017,Merolla2014_truenorth}. 

\begin{figure}
\begin{center}
\includegraphics[scale=0.65]{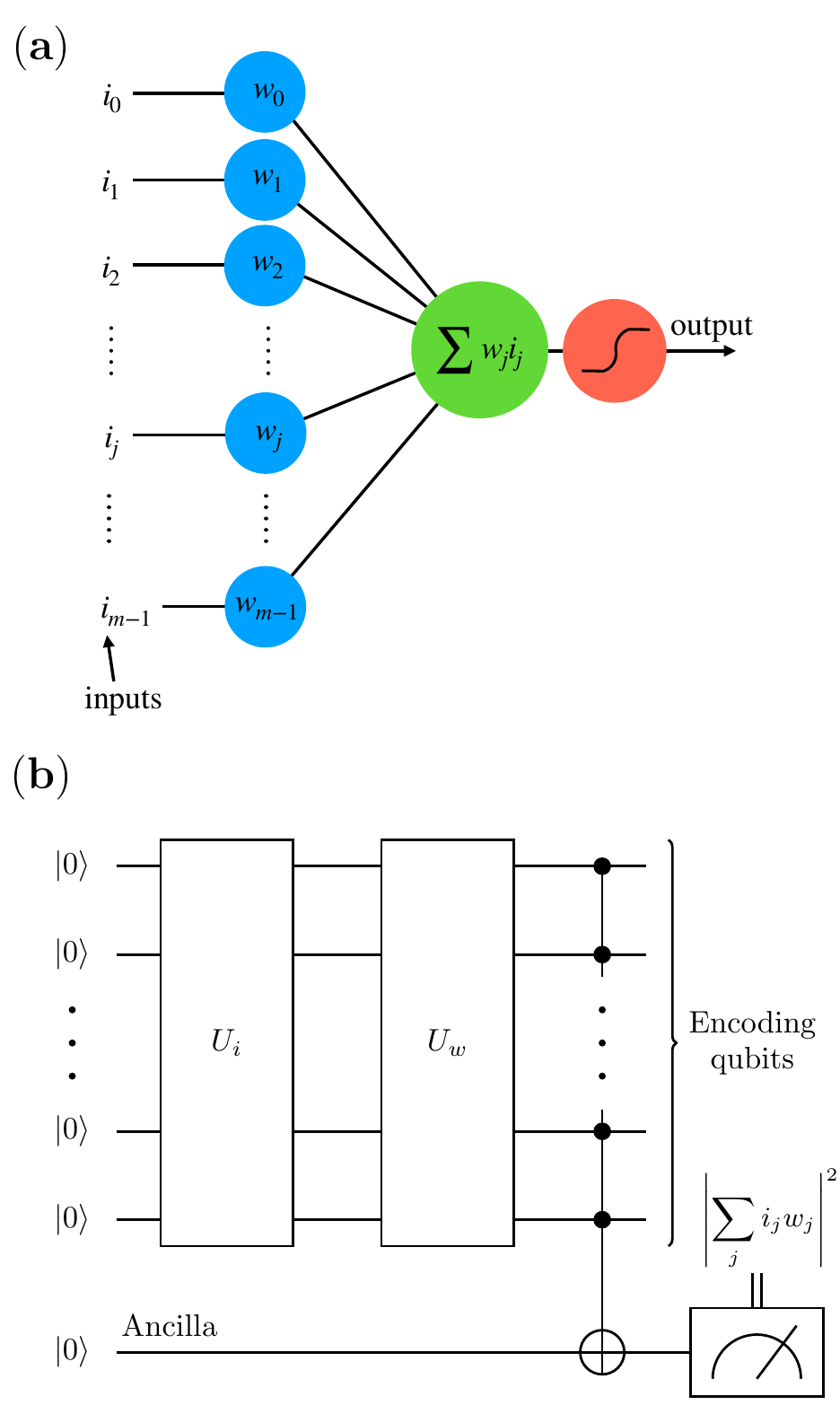}
\caption{{\bf Perceptron models.} (a) Schematic outline of the classical perceptron as a model of artificial neuron: An input array $\vec{i}$ is processed with a weight vector $\vec{w}$ to produce a linear, binary valued output function. In its simplest realization, also the elements of $\vec{i}$ and $\vec{w}$ are binary valued, the perceptron acting as a binary (linear) classifier. (b) Scheme of the quantum algorithm for the implementation of the artificial neuron model on a quantum processor: From the system initialized in its idle configuration, the first two unitary operations prepare the input quantum state, $|\psi_i\rangle$, and implement the $U_w$ transformation, respectively. The final outcome is then written on an ancilla qubit, which is eventually measured to evaluate the activation state of the perceptron.}
\label{figperceptron_classic}
\end{center}
\end{figure}

Among the possible computing platforms, prospective quantum computers seem particularly well suited for implementing  artificial neural networks \cite{Biamonte2017Nature}. 
In fact, the intrinsic property of Quantum Mechanics of representing and storing large complex valued vectors and matrices, as well as performing linear operations on such vectors, is believed to result in an exponential increase either in memory storage or processing power for neural networks directly implemented on quantum processors \cite{Neukart_QCANN_2013,Schuld_Petruccione_review_2014,Schuld2015PhysLettA,Kapoor2016NIPS,Lloyd_quantum_algorithms_machine_learning_arxiv_2016,Schuld2017EPL,Lamata2017SciRep,Alvarez2017SciRep,Otterbach2017,Rebentrost2018PRA}.
The simplest model of an artificial neuron, the so called ``perceptron", was originally proposed by R. Rosenblatt in 1957 \cite{Rosenblatt1957}, and is schematically outlined in Fig.~\ref{figperceptron_classic}(a). A real valued vector, $\vec{i}$, of dimension $m$ represents the input, and it is combined with a real valued weight vector, $\vec{w}$. The perceptron output is evaluated as a binary response function resulting from the inner product of the two vectors, with a threshold value deciding for the ``yes/no" response. In the lowest level implementations, $\vec{i}$ and $\vec{w}$ are binary valued vectors themselves, as proposed by McCulloch and Pitts in 1943 as a simple model of a neuron~\cite{McCulloch_Pitts_1943,Zurada:intro_ANN_1992}. 

Perceptrons and McCulloch-Pitts neurons are limited in the operations that they can perform, but they are still at the basis of machine learning algorithms in more complex artificial neural networks in multilayered perceptron architectures. However, the computational complexity increases with increasing number of nodes and interlayer connectivity and it is not clear whether this could eventually call for a change in paradigm, although different strategies can be put forward to optimize the efficiency of classical algorithms \cite{Mocanu2018ncom}.
In this respect, several proposals have been advanced in recent years to implement perceptrons on quantum computers. The most largely investigated concept is that of a ``qubit neuron", in which each qubit (the computational unit in quantum computers) acts as an individual neuron within the network. Most of the research effort has been devoted to exploit the nonlinearity of the measurement process in order to implement the threshold function \cite{Schuld_Petruccione_review_2014}. 

Here we introduce an alternative design that closely mimics a Rosenblatt perceptron on a quantum computer. First, the equivalent of $m$-dimensional classical input and weight vectors is encoded on the quantum hardware by using $N$ qubits, where $m=2^N$. On one hand, this evidently allows to exploit the exponential advantage of quantum information storage, as already pointed out \cite{Schuld2015PhysLettA,Schuld2017EPL}. On the other, we implement an original procedure to generate multipartite entangled states based on quantum information principles \cite{Rossi2013} that allows to crucially scale down the quantum computational resources to be employed. We experimentally show the effectiveness of such an approach by practically implementing a 2 qubits version of the algorithm on the IBM quantum processor available for cloud quantum computing. In this respect, the present work constitutes a key step towards the efficient use of prospective quantum processing devices for machine learning applications. Remarkably, we show that the quantum perceptron model can {be used to} sort out simple patterns, such as vertical or horizontal lines among all possible inputs. In order to show the potential of our proposed implementation of a quantum artificial neuron, we theoretically simulate a 4+1 qubits version using the IBM quantum simulator. We conclude the paper by discussing the usefulness of our algorithm as a quantum neuron in fully quantum neural networks.

\section{Quantum circuit modeling of a classical perceptron}
\label{Outline of the quantum perceptron}

A scheme of the quantum algorithm proposed in this work is shown in Fig.~\ref{figperceptron_classic}(b). The input and weight vectors are limited to binary values, $i_j,w_j\in \{-1,1\} $, as in McCulloch-Pitts neurons. Hence, a $m$-dimensional input vector is encoded using the $m$ coefficients needed to define a general wavefunction $|\psi_i\rangle$ of $N$ qubits. In practice, given arbitrary input ($\vec{i}$) and weight ($\vec{w}$) vectors
\begin{equation}
\vec{i} = \begin{pmatrix}
    i_{0} \\
    i_{1} \\
    \vdots \\
    i_{m-1}
\end{pmatrix},\,\,\,
\vec{w} = \begin{pmatrix}
    w_{0} \\
    w_{1} \\
    \vdots \\
    w_{m-1}
\end{pmatrix}
\end{equation}
with $i_j,w_j \in \{-1,1\}$, we define the two quantum states
\begin{equation}
|\psi_i\rangle = \frac{1}{\sqrt{m}}\sum_{j = 0}^{m - 1} i_j |j\rangle  ; \,\,\,
|\psi_w\rangle = \frac{1}{\sqrt{m}}\sum_{j = 0}^{m - 1} w_j |j\rangle   \,\, .
\label{eq:inputstate}
\end{equation}
The states $|j\rangle \in \{|00\ldots 00\rangle ,|00\ldots 01\rangle ,\ldots,|11\dots 11\rangle\}$ form the so called computational basis of the quantum processor, i.e.\ the basis in the Hilbert space of $N$ qubits, corresponding to all possible states of the single qubits being either in $|0\rangle$ or $|1\rangle$. As usual, these states are labeled with integers $j\in\{0,\ldots,m-1\}$ arising from the decimal representation of the respective binary string. Evidently, if $N$ qubits are used in the register, there are $m=2^N$ basis states labelled $|j\rangle$ and, as outlined in Eq.~\eqref{eq:inputstate}, we can use factors $\pm 1$ to encode the $m$-dimensional classical vectors into an uniformly weighted superposition of the full computational basis.

The first step of the algorithm prepares the state $|\psi_i\rangle$ by encoding the input values in $\vec{i}$. Assuming the qubits to be initialized in the state $|00\ldots00\rangle \equiv |0\rangle^{\otimes N}$, we perform a unitary transformation $U_i$ such that
\begin{equation}
U_i|0\rangle^{\otimes N}=|\psi_i\rangle \,\, .
\end{equation}
In principle, any $m\times m$ unitary matrix having $\vec{i}$ in the first column can be used to this purpose, and we will give explicit examples in the following. Notice that, in a more general scenario, the preparation of the input state starting from a blank register might be replaced by a direct call to a quantum memory \cite{Giovannetti2008PRL} where $|\psi_i\rangle$ was previously stored.

The second step computes the inner product between $\vec{w}$ and $\vec{i}$ using the quantum register.
This task can be performed efficiently by defining a unitary transformation, $U_w$, such that the weight quantum state is rotated as
\begin{equation}
U_w |\psi_w\rangle = |1\rangle^{\otimes N} = |m-1\rangle \,\, .
\label{eq:UwConstraint}
\end{equation}
As before, any $m\times m$ unitary matrix having $\vec{w}^T$ in the last row satisfies this condition. If we apply $U_w$ after $U_i$, the overall $N$-qubits quantum state becomes
\begin{equation}
U_w |\psi_i\rangle = \sum_{j = 0}^{m - 1} c_j |j\rangle \equiv |\phi_{i,w}\rangle \, .
\label{eq:afterUs}
\end{equation}
Using Eq.~\eqref{eq:UwConstraint}, the scalar product between the two quantum states is
\begin{equation}
\begin{aligned}
\langle \psi_w | \psi_i\rangle & = \langle \psi_w | U_w^\dagger U_w | \psi_i\rangle = \\
& = \langle m-1 |\phi_{i,w}\rangle = c_{m-1} \, ,
\end{aligned}
\label{eq:idotw}
\end{equation}
and from the definitions in Eq.~\eqref{eq:inputstate} it is easily seen that the scalar product of input and weight vectors is $\vec{w}\cdot\vec{i} = m\langle \psi_w | \psi_i\rangle$. 
Therefore, the desired result is contained, up to a normalization factor, in the coefficient $c_{m-1}$ of the final state $|\phi_{i,w}\rangle$.

In order to extract such an information, we propose to use an ancilla qubit ($a$) initially set in the state $|0\rangle$.
A multi-controlled $\mathrm{NOT}$ gate between the $N$ encoding qubits and the target $a$ leads to \cite{NielsenChuang}:
\begin{equation}
|\phi_{i,w}\rangle|0\rangle_a \rightarrow \sum_{j = 0}^{m - 2} c_j |j\rangle|0\rangle_a + c_{m-1}|m-1\rangle|1\rangle_a
\end{equation}

The nonlinearity required by the threshold function at the output of the perceptron is immediately obtained by performing a quantum measurement: indeed, by measuring the state of the ancilla qubit in the computational basis produces the output $|1\rangle_a$ (i.e., an activated perceptron) with probability $|c_{m-1}|^2$. As it will be shown in the following, this choice proves simultaneously very simple and effective in producing the correct result. 
However, it should be noticed that refined threshold functions can be applied once the inner product information is stored on the ancilla \cite{Hu2018_Toward_real_quantum_neuron,Cao_2017_Quantum_neuron_building_block,Torrontegui_2018_Quantum_perceptron_unitary_approximator}.  
We also notice that both parallel and anti-parallel $\vec{i}$-$\vec{w}$ vectors produce an activation of the perceptron, while orthogonal vectors always result in the ancilla being measured in the state $|0\rangle_a$. 
This is a direct consequence of the probability being a quadratic function, i.e. $|c_{m-1}|^2$ in the present case, at difference with classical perceptrons that can only be employed as linear classifiers in their simplest realizations. 
In fact, our quantum perceptron model can be efficiently used as a pattern classifier, as it will be shown below, since it allows to interpret a given pattern and its negative on equivalent footing. 
Formally, this intrinsic symmetry reflects the invariance of the encoding $|\psi_i\rangle$ and $|\psi_w\rangle$ states under addition of a global $-1$ factor.

\section{Implementation of the unitary transformations}
\label{Implementation of the two unitary transformations}

\begin{figure*}
\centering
\includegraphics[scale=1]{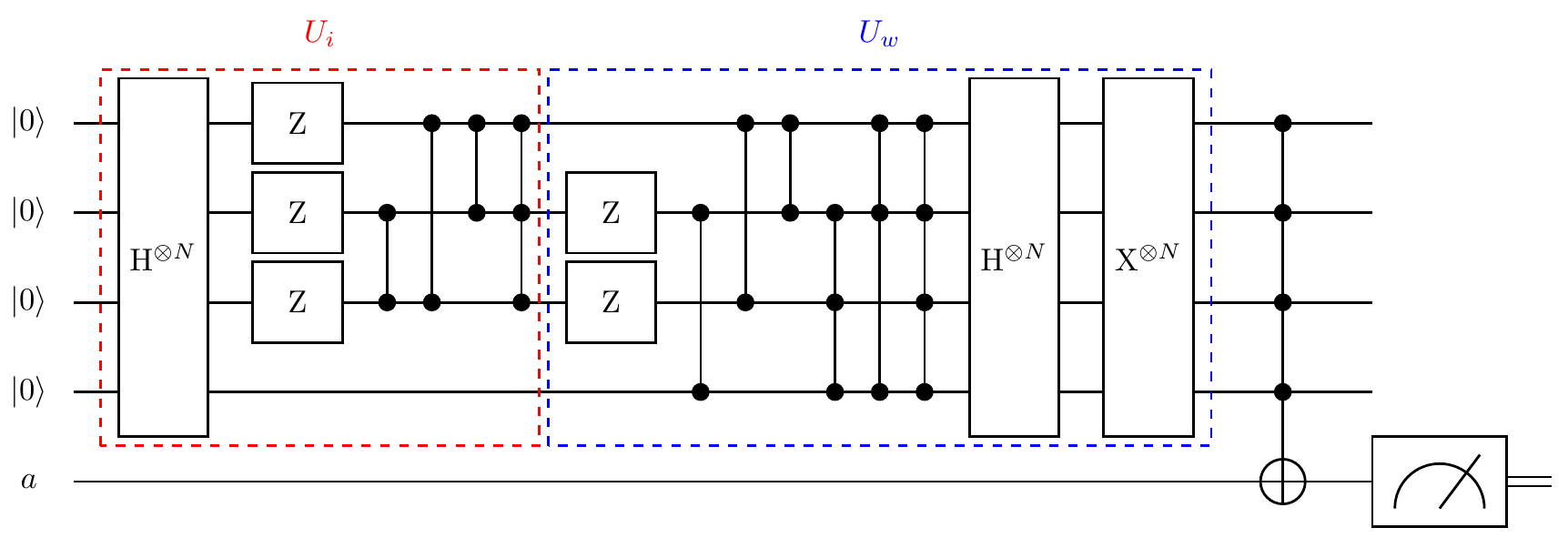} 
\caption{{\bf Quantum circuit of a $N=4$ perceptron.} An example of a typical quantum circuit for a perceptron model with $N=4$ qubits (i.e. capable of processing $m=2^4=16$ dimensional input vectors), which employs the algorithm for the generation of hypergraph states, including the HSGS (see main text). In this example, the input vector has elements $i_0 = i_1 = -1$, and $i_j = 1$ for $j=2,\dots,15$, while the weight vector has elements $w_2 = w_3 = w_4 = -1$, and $1$ in all other entries. Multi-controlled $\mathrm{C}^p\mathrm{Z}$ gates are denoted by vertical lines and black dots on the qubits involved. The HSGS is realized inside the $U_i$ block after the initial $\mathrm{H}^{\otimes N}$ gate, and in the $U_w$ block before the final $\mathrm{H}^{\otimes N}$ and $\mathrm{NOT}^{\otimes N}$ operations.}
\label{fig:Sequence}
\end{figure*}

One of the most critical tasks to be practically solved when implementing a quantum neural network model is the efficient implementation of unitary transformations. In machine learning applications, this might eventually discriminate between algorithms that show truly quantum advantage over their classical counterparts  \cite{Schuld2017EPL}. 
Here we discuss an original strategy for practically implementing  the preparation of the input state $|\psi_i\rangle$ and the unitary transformation $U_w$ on a quantum hardware. 
In particular, we will first outline the most straightforward algorithm one might think of employing, i.e. the ``brute force'' application of successive sign flip blocks. 
Then, we will show an alternative and more effective approach based on the generation of hypergraph states. 
In the next Section we will see that only the latter allows to practically implement this quantum perceptron model on a real quantum device. 

So, as a first step we define a sign flip block, $\mathrm{SF}_{N,j}$, as the unitary transformation acting on the computational basis of $N$ qubits in the following way:
\begin{equation}
\mathrm{SF}_{N,j} |j'\rangle = \begin{cases} |j'\rangle \quad & \text{if } j \neq j' \\ -|j'\rangle \quad & \text{if } j = j'\end{cases} \, .
\label{eq:SFdef}
\end{equation}
For any $N,m=2^N$, a controlled $\mathrm{Z}$ operation between $N$ qubits ($\mathrm{C}^N\mathrm{Z}$) is a well known quantum gate~\cite{NielsenChuang} realizing $\mathrm{SF}_{N,m-1}$, while a single qubit $\mathrm{Z}$ gate acts as $\mathrm{SF}_{1,1}$. We can therefore implement in practice the whole family of sign-flip blocks for $N$ qubits by using $\mathrm{C}^N\mathrm{Z}$ gates in combination with single qubit $\mathrm{NOT}$ gates (i.e.\ single bit flip operations):
\begin{equation}
\mathrm{SF}_{N,j} = \mathrm{O}_j\left(\mathrm{C}^N\mathrm{Z}\right)\mathrm{O}_j \, ,
\end{equation}
where
\begin{equation}
\mathrm{O}_j = \bigotimes_{l=0}^{m-1} (\mathrm{NOT}_l)^{1-j_l} \, .
\end{equation}
In the expression above, $\mathrm{NOT}_l$ means that the bit flip is applied to the $l$-th qubit and $j_l = 0$($1$) if the $l$-th qubit is in state $|0\rangle$($|1\rangle$) in the computational basis state $| j \rangle$. 
Alternatively, the same result can also be obtained by using an extra ancillary qubit and multi-controlled $\mathrm{NOT}$ gates ($\mathrm{C}^N\mathrm{NOT}$), i.e.\ bit flip operations conditioned on the state of some control qubits. 
We explicitly point out that, as it is easily understood from the definition in Eq.~\eqref{eq:SFdef}, any $\mathrm{SF}_{N,j}$ is the inverse of itself. 
Then, the full sequence to implement $U_i$ can be summarized as follows: starting from the initialized register $|0\rangle^{\otimes N}$, parallel Hadamard ($\mathrm{H}$) gates are applied to create an equal superposition of all the elements of the computational basis:
\begin{equation}
|0\rangle^{\otimes N} \xrightarrow[]{\mathrm{H}^{\otimes N}} \frac{1}{\sqrt{m}}\sum_{j = 0}^{m - 1} |j\rangle \equiv |\psi_0\rangle \, ,
\label{eq:initialh}
\end{equation}
where we remind that \cite{NielsenChuang}
\begin{equation}
\mathrm{H} |0\rangle  = \frac{|0\rangle+|1\rangle}{\sqrt{2}}   \,\,;  \,\,\,
\mathrm{H} |1\rangle  = \frac{|0\rangle-|1\rangle}{\sqrt{2}} \,\,.
\end{equation}
Then, the $\mathrm{SF}_{N,j}$ blocks are applied one by one whenever there is a $-1$ factor in front of $|j\rangle$, in the representation of the target $|\psi_i\rangle$. Notice that any $\mathrm{SF}_{N,j}$ only affects a single element of the computational basis while leaving all others unchanged. Moreover, all $\mathrm{SF}_{N,j}$ blocks commute with each other, so they can actually be performed in any order. As already anticipated, the whole problem is symmetric under the addition of a global $-1$ factor (i.e.\ $|\psi_i\rangle$ and $-|\psi_i\rangle$ are fully equivalent). Hence, there can be only at most $m/2 = 2^{N-1}$ independent $-1$ factors, and $2^{N-1}$ sign flip blocks are needed in the worst case. 
A similar strategy can also be applied to implement the other unitary operation in the quantum perceptron algorithm, $U_w$. 
Indeed, applying first the $\mathrm{SF}_{N,j}$ blocks that would be needed to flip all the $-1$ signs in front of the computational basis elements in the associated $|\psi_w\rangle$ leads to the balanced superposition $|\psi_w\rangle \rightarrow |\psi_0\rangle$. This quantum state can then be brought into the desired $|11\dots11\rangle \equiv |1\rangle^{\otimes N}$ state by applying parallel Hadamard and $\mathrm{NOT}$ gates:
\begin{equation}
|\psi_0\rangle \xrightarrow[]{\mathrm{H}^{\otimes N}} |0\rangle^{\otimes N} \xrightarrow[]{\mathrm{NOT}^{\otimes N}} |1\rangle^{\otimes N} \, .
\label{eq:HnNn}
\end{equation}

\begin{figure*}
\centering
\includegraphics[scale=0.8]{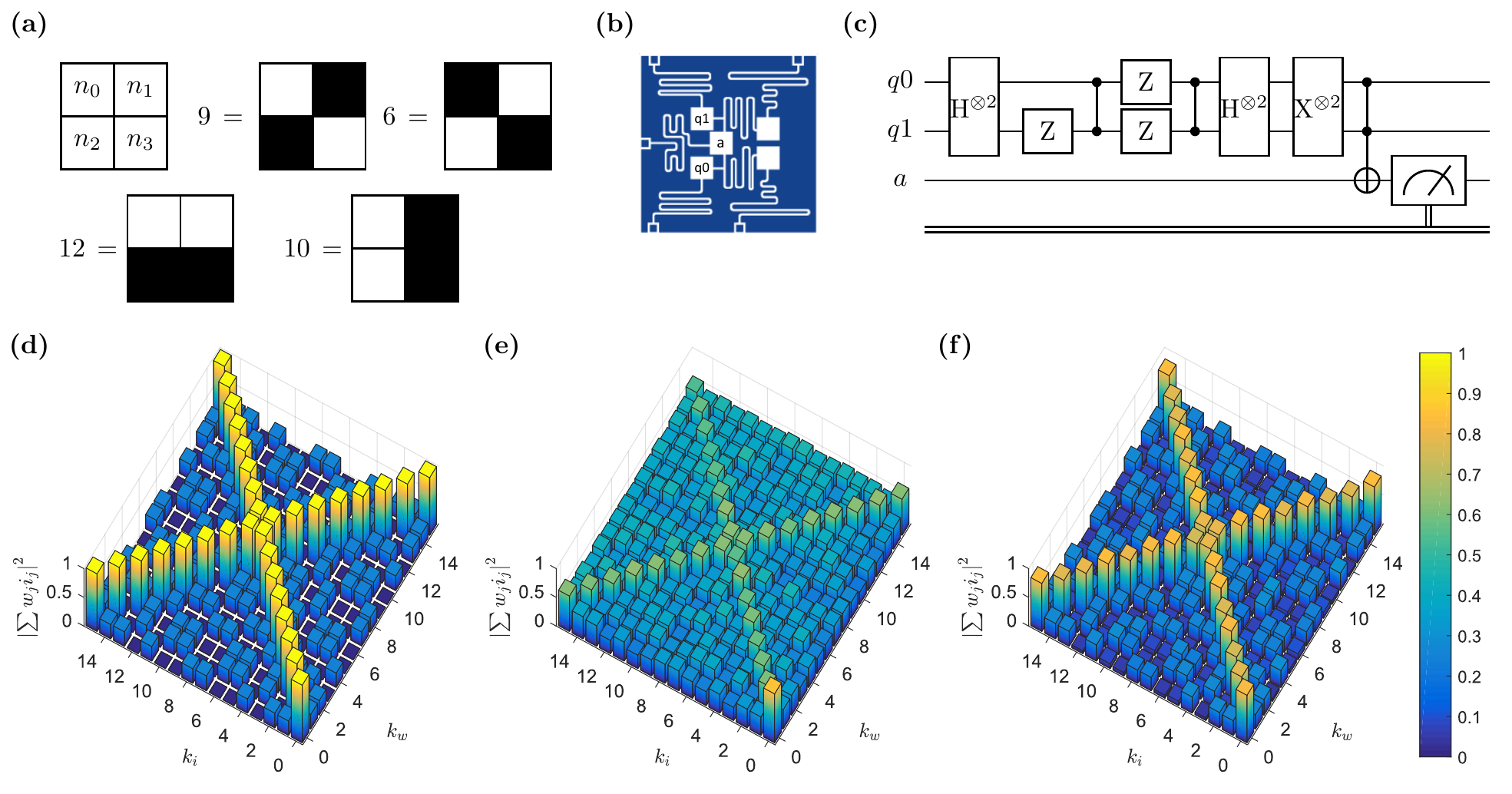} 
\caption{{\bf Results for $N = 2$ quantum perceptron model.}  (a) Scheme used to label the 2$\times$2 patterns and a few examples of patterns. (b) Scheme of IBM Q-5 ``Tenerife'' backend quantum processor. (c) Example of the gate sequence for the $N=2$ case, with input and weight vectors corresponding to labels $k_i=11$ and $k_w=7$. (d) Ideal outcome of the quantum perceptron algorithm, simulated on a classical computer. (e) Results from the Tenerife processor using the algorithm with multi-controlled sign flip blocks. (f) Results from the Tenerife processor using the algorithm for the generation of hypergraph states.}
\label{fig:ResultsIbmqx4}
\end{figure*}

Evidently, the above strategy is exponentially expensive in terms of circuit depth as a function of the number of qubits, and requires an exponential number of $N$-controlled quantum gates.

On the other hand, a more efficient strategy can be given after realizing that the class of possible input- and weight-encoding states, Eq.~\eqref{eq:inputstate}, coincides with the set of the so called {\it hypergraph} states. The latter are ubiquitous ingredients of many renown quantum algorithms, and have been extensively studied and theoretically characterized~\cite{Rossi2013,Ghio2017}. 
In particular, hypergraph states can be mapped into the vertices and hyper-edges of generalized graphs, and can be prepared by using single qubit and (multi)-controlled $\mathrm{Z}$ gates, with at most a single $N$-controlled $\mathrm{C}^N\mathrm{Z}$ and with the possibility of performing many $p$-controlled $\mathrm{C}^p\mathrm{Z}$ gates (involving only $p$ qubits, with $p < N$) in parallel. 
After an initial $\mathrm{H}^{\otimes N}$ gate (see Eq.~\eqref{eq:initialh}), the algorithm takes a series of iterative steps~\cite{Rossi2013} that are described below. 
In the following, we will refer to this portion of the algorithm to generate hypergraph states as the ``hypergraph states generation subroutine'' (HSGS).  \\
First, we check whether there is any component with only one qubit in state $|1\rangle$ (i.e.\ of the form $|0\ldots010\ldots0\rangle$) requiring a $-1$ factor, in the representation of $|\psi_i\rangle$ on the computational basis. If so, the corresponding single qubit $\mathrm{Z}$ gate is applied by targeting the only qubit in state $|1\rangle$. Notice that this might introduce additional $-1$ factors in front of states with more than one qubit in state $|1\rangle$. Then, for $p = 2,\ldots,N$, we consider the components of the computational basis with exactly $p$ qubits in state $|1\rangle$. For each of them, an additional $-1$ sign is introduced in front of its current amplitude (if it is needed and it was not previously introduced) by applying the corresponding $\mathrm{C}^p\mathrm{Z}$ between the $p$ qubits in state $|1\rangle$. 
Similarly, if an unwanted sign is already present due to a previous step, this can be easily removed by applying the same $\mathrm{C}^p\mathrm{Z}$. Since $\mathrm{C}^p\mathrm{Z}$ acts non trivially only on the manifold with $p$ or more qubits being in state $|1\rangle$, the signs of all the elements with a lower number of $|1\rangle$ components are left unchanged. 
As a consequence, when $p=N$ all the signs are the desired ones. As in the previous case, $U_w$ can be obtained by slightly modifying the sequence of gates that would be used to generate $|\psi_w\rangle$. Indeed, one can start by first performing the HSGS tailored according to the $\pm 1$ factors in $|\psi_w\rangle$. Since all the gates involved in HSGS are the inverse of themselves and commute with each other, this step is equivalent to the unitary transformation bringing $|\psi_w\rangle$ back to the equally balanced superposition of the computational basis states $|\psi_0\rangle$. The desired transformation $U_w$ is finally completed by adding parallel $\mathrm{H}^{\otimes N}$ and $\mathrm{NOT}^{\otimes N}$ gates (see Eq.~\eqref{eq:HnNn}). 
An example of the full sequence for a specific $N = 4$ case is shown, e.g., in Fig.~\ref{fig:Sequence}. Notice that our optimized algorithm involving hypergraph states successfully reduces the required quantum resources with respect to a brute force approach, even if it still involves an exponential cost in terms of circuit depth or clock cycles on the quantum processor in the worst case.

Before proceeding, it is probably worth pointing out the role of $U_w$ in this algorithm, which is essentially to cancel some of the transformations performed to prepare $|\psi_i\rangle$, or even all of them if the condition $\vec{i}=\vec{w}$ is satisfied. 
Further optimization of the algorithm, lying beyond the scope of the present work, might therefore be pursued at the compiling stage. However, notice that the input and weight vectors can, in practical applications, remain unknown or hidden until runtime.


\section{Numerical results and Quantum simulations}
\label{Results}

We implemented the algorithm for a single quantum perceptron both on classical simulators working out the matrix algebra of the circuit and on cloud-based quantum simulators, specifically the IBM Quantum Experience real backends~\cite{Note1}, using the Qiskit Python development kit~\cite{Note2}. Due to the constraints imposed by the actual IBM hardware in terms of connectivity between the different qubits, we limited the real quantum simulation  to the $N=2$ case. Nevertheless, even this small-scale example is already sufficient to show all the distinctive features of our proposed set up, such as the exponential growth of the analyzable problems dimension, as well as the pattern recognition potential. In general, as already mentioned, in this encoding scheme $N$ qubits can store and process $2^N$-dimensional input and weight vectors, and thus $2^{2^N}$ different input patterns can be analyzed against the same number of the different $\vec{w}$ that are possible. Moreover, all binary inputs and weights can easily be converted into black and white patterns, thus providing a visual interpretation of the activity of the artificial neuron.

Going back to the case study with $N=2$, $2^2 = 4$ binary images can be managed, and thus $2^{2^2} = 16$ different patterns could be analyzed. The conversion between $\vec{i}$ or $\vec{w}$ and $2\times 2$ pixels visual patterns is done as follows. As depicted in Fig.~\ref{fig:ResultsIbmqx4}a, we label each image ordering the pixels left to right, top to bottom, and assigning a value $n_j = 0$($1$) to a white (black) pixel. The corresponding input or weight vector is then built by setting $i_j=(-1)^{n_j}$ (or $w_j=(-1)^{n_j}$). We can also univocally assign an integer label $k_i$ (or $k_w$) to any pattern by converting the binary string $\mathtt{n}_0\mathtt{n}_1\mathtt{n}_2\mathtt{n}_3$ to its corresponding decimal number representation. 
Under this encoding scheme, e.g., numbers 3 and 12 are used to label patterns with horizontal lines, while 5 and 10 denote patterns with vertical lines, and 6 and 9 are used to label images with checkerboard-like pattern. 
An example of the sequence of operations performed on the IBM quantum computer using hypergraph states is shown in Fig.~\ref{fig:ResultsIbmqx4}c for $\vec{i}$ corresponding to the index $k_i=11$, and $\vec{w}$ corresponding to $k_w=7$. 

The Hilbert space of 2 qubits is relatively small, with a total of 16 possible values for $\vec{i}$ and $\vec{w}$. Hence, the quantum perceptron model could be experimentally tested on the IBM quantum computer for all possible combinations of input and weights. 
The results of these experiments, and the comparison with classical numerical simulations, are shown in Fig.~\ref{fig:ResultsIbmqx4}d-f. First, we plot the ideal outcome of the quantum perceptron algorithm in Fig.~\ref{fig:ResultsIbmqx4}d, where both the global $-1$ factor and the input-weight symmetries are immediately evident. In particular, for any given weight vector $\vec{w}$, the perceptron is able to single out from the $16$ possible input patterns only $\vec{i}=\vec{w}$ and its negative (with output $|c_{m-1}|^2 = 1$, i.e.\ the perfect activation of the neuron), while all other inputs give outputs smaller than $0.25$. If the inputs and weights are translated into $2\times 2$ black and white pixel grids, it is not difficult to see that a single quantum perceptron can be used to recognize, e.g.,\ vertical lines, horizontal lines, or checkerboard patterns.

The actual experimental results are then shown in Fig.~\ref{fig:ResultsIbmqx4}e-f, where the same algorithm is run on the IBM Q 5 ``Tenerife" quantum processor \cite{Note3}. 
First, we show in panel~\ref{fig:ResultsIbmqx4}e the results of the first, non-optimized approach introduced in the previous Section, which makes direct use of sign flip blocks. We deliberately did not take into account the global sign symmetry, thus treating any \ $|\psi_i\rangle$ and $-|\psi_i\rangle$ as distinct input quantum states and using up to $2^N$ sign flip blocks. We notice that even in such an elementary example the algorithm performs worse and worse with increasing number of blocks. Notice, however, that despite the quantitative inaccuracy of the quantum simulated outputs, the underlying structure of the output is already quite clear. \\
On the other hand, a remarkably better accuracy, also on the quantitative side and with small errors, is obtained when using the algorithm based on the hypergraph states formalism, whose experimental results are shown in panel~\ref{fig:ResultsIbmqx4}f and represent the main result of this work. In this case, the global phase symmetry is naturally embedded in the algorithm itself, and the results show symmetric performances all over the range of possible inputs and weights. All combinations of $\vec{i}$ and $\vec{w}$ yield results either larger than 0.75 or smaller than 0.3, in very good quantitative agreement with the expected results plotted in panel~\ref{fig:ResultsIbmqx4}d. As a technical warning, we finally notice that in all of the three cases shown in panels d-f of Fig.~\ref{fig:ResultsIbmqx4}, the $\mathrm{C}^pZ$ operations were obtained by adding single qubit Hadamard gates on the target qubit before and after the corresponding $\mathrm{C}^p\mathrm{NOT}$ gate. For $p=1$ this is a $\mathrm{CNOT}$ gate, which is natively implemented on the IBM quantum hardware, while the case $p=2$ is known as the Toffoli gate, for which a standard decomposition into 6 $\mathrm{CNOT}$s and single qubit rotations is known~\cite{NielsenChuang}. 

\begin{figure}
\centering
\includegraphics[scale=0.8]{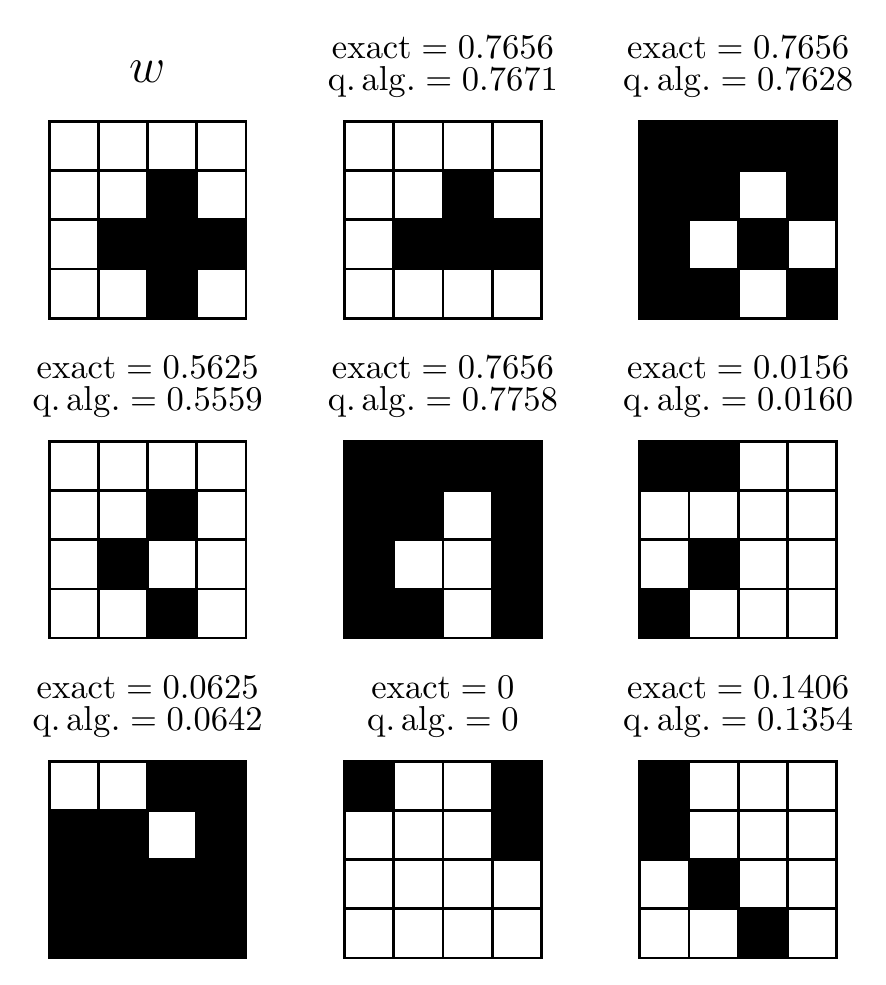} 
\caption{{\bf Pattern recognition for $N = 4$.} A possible choice of the weight vector for the $N = 4$ case is represented in the first panel (top left), and a small selection of  different input vectors are then simulated with the quantum perceptron model. Above each input pattern, the quantitative answers of the artificial neuron are reported, as obtained either through standard linear algebra (ideal results) or resulting from the simulation of the quantum algorithm (run on a classical computer).}
\label{fig:ResultsSIMN4}
\end{figure}

Finally, in the spirit of showing the potential scalability and usefulness of this quantum perceptron model for classification purposes, we have applied the algorithm to the $N=4$ qubits case by using the circuit simulator feature available in Qiskit \cite{Note4}. Now, there are a total $2^{32}$ possible combinations of $\vec{i}$ and $\vec{w}$ vectors, far too many to explore the whole combinatorial space as previously done for the 2 qubits in Fig.~\ref{fig:ResultsIbmqx4}. 
To explicitly show a few examples,  we have chosen a single weight vector corresponding to a simple cross-shaped pattern when represented as a 4$\times$4 pixels image (encoded along the same lines of the $N=2$ case, see first panel in Fig.~\ref{fig:ResultsIbmqx4}), and weighted it against several possible choices of input vectors. Some results are reported in Fig.~\ref{fig:ResultsSIMN4} for a selected choice of input vectors, where the artificial neuron output is computed both with standard linear algebra and simulated with a quantum circuit on a virtual quantum simulator run on a classical computer. 
Evidently, there is an overall excellent agreement when comparing the two values for each pattern, within statistical inaccuracy due to the finite number ($n_{shots} = 8192$) of repetitions imposed on the quantum sequence used to estimate the probability $|c_{m-1}|^2$. The perceptron is able to discriminate the weight pattern (and its negative) giving an output larger than 0.5 for all images that differ from the weight or its negative by 2 bits or less.

\section{Conclusions and discussion}
\label{Conclusions}


In summary, we have proposed a model for perceptrons to be directly implemented on near-term quantum processing devices, and we have experimentally tested it on a 5-qubits IBM quantum computer based on superconducting technology. 
Our algorithm presents an exponential advantage over classical perceptron models, as we have explicitly shown by representing and classifying 4 bits strings using 2 qubits, and 16 bits strings using only 4 qubits. 

The problem of exponential advantage requires a separate discussion. In principle, generic quantum states or unitary transformations require an exponentially large number of elementary gates to be implemented, and this could somehow hinder the effective advantages brought by quantum computers for machine learning applications. This currently represents a general problem of most quantum machine learning algorithms. Moreover, with increasing $N$, severe issues can arise from the practical necessity to decompose multiply controlled operations by only using single- and two-qubit gates \cite{Bergholm:2005:state_preparation_decomposition,Plesch:2011:state_preparation_decomposition_better}. However, this limitation strictly depends on the effective constraints imposed by the given quantum processor and on the required degree of accuracy. In fact, it has been shown that several classes of quantum states can be approximated efficiently with arbitrary precision, with oracle based approaches \cite{Grover2002,Soklakov2006,Clader2013} or by using a number of multi-controlled rotations that is linear with the number of qubits \cite{Mosca:01:state_preparation_multicontrolled}. Using these results, it might then be possible to design a version of our proposed quantum perceptron algorithm working with approximated encoding quantum states instead of exact ones, which would have the potential to scale exponentially better than any classical algorithm implementing a perceptron model. In this respect, it is also worth pointing out that our procedure is fully general and could be implemented and run on any platform capable of performing universal quantum computation. While we have employed a quantum hardware that is based on superconducting technology and qubits, a very promising alternative is the trapped-ion based quantum computer \cite{Schindler2013Njp}, in which multi-qubit entangling gates might be readily available \cite{Molmer1999PRL,Schindler2013NPhys}.

As a further strategy for future developments, we notice that in the present work we restricted the whole analysis to binary inputs and weight vectors (the so called ``McCollough-Pitts'' neuron model), mainly for clarity and simplicity of implementation. A possible improvement for the algorithm presented is obviously to encode continuously valued vectors (equivalent to grey scale images). This could be achieved by using continuously valued phase factors in $|\psi_i\rangle$ and $|\psi_w\rangle$ \cite{Schuld2015PhysLettA}. Finally, a potentially very exciting continuation of this work would be to connect multiple layers of our quantum perceptrons to build a feedforward deep neural network, which could be fully run on dedicated quantum hardware. In such a network, each neuron could use two ancilla qubits, one to be measured to introduce the nonlinearity as done in this work, while the second would be used to propagate the information from each neuron to the successive layer in the network in a fully quantum coherent way. As such, our work thus constitute a concrete first step towards an actual application of near-term (i.e., with few tens of non-error corrected qubits) quantum processors to be employed as fast and efficient trained artificial quantum neural networks.

\section{Aknowledgements}
We acknowledge the University of Pavia Blue Sky Research project number BSR1732907. 
This research was also supported by the Italian Ministry of Education, University and Research (MIUR): ``Dipartimenti di Eccellenza Program (2018-2022)", Department of Physics, University of Pavia. 
{We acknowledge use of the IBM Quantum Experience for this work. The views expressed are those of the authors and do not reflect the official policy or position of IBM company or the IBM-Q team.}

\end{document}